# Possible nucleus of the Bergman cluster in the Zn-Mg-Y alloy system

Kei Nakayama[a*], Masaya Nakagawa[a] and Yasumasa Koyama[ab]

[a]*Department of Electronic and Physical Systems, Waseda University, 3-4-1, Okubo, Shinjuku-ku, Tokyo 169-8555, Japan*

[b]*Kagami Memorial Research Institute for Materials Science and Technology, Waseda University, 2-8-26, Nishiwaseda, Shinjuku-ku, Tokyo 169-0051, Japan*

*Corresponding author. Email: d.i.y.999@fuji.waseda.jp

# Possible nucleus of the Bergman cluster in the Zn-Mg-Y alloy system

To understand the formation of the Bergman cluster in the F-type icosahedral quasicrystal, crystallographic relations between the quasicrystal and the intermetallic-compound H and $Zn_{23}Y_6$ phases in the Zn-Mg-Y alloy system were investigated mainly by transmission electron microscopy. It was found that, although sample rotations of about 1° were required to obtain simple crystallographic relations, the orientation relationship was established among the cubic-$Fm\bar{3}m$ $Zn_{23}Y_6$ structure, the hexagonal-$P6_3/mmc$ H structure, and the F-type icosahedral quasicrystal (IQ); that is, $[\bar{1}13]_c$ // the fivefold axis in the IQ // $N(2\bar{1}\bar{1}0)_H$, and $[110]_c$ // the twofold axis in the IQ // $N(05\bar{5}3)_H$, where $N(hkml)_H$ means the normal direction of the $(hkml)_H$ plane in the H structure. The correspondences between atomic positions in the Bergman cluster and in the $Zn_{23}Y_6$ structure and between those in the cluster and in the H structure were investigated on the basis of the established relationship. As a result, an assembly of six short-penetrated-decagonal columns was identified as an appropriate nucleus in the formation of the Bergman cluster from these two structures.

## I. Introduction

A three-dimensional icosahedral quasicrystal was discovered in the Al-Mn alloy system by Schechtman and coworkers [1]. From studies on quasicrystals since then, icosahedral quasicrystals are known to be characterized by the presence of both quasiperiodicity and a fivefold rotational axis [2–4]. As for their atomic arrangements, icosahedral quasicrystals involve giant atomic clusters with some coordination shells, and Bergman, Mackay, and Tsai clusters are three typical examples of such clusters [5–7]. Among these clusters, the Bergman cluster was found in both the icosahedral quasicrystal and its approximant-T phase in the Zn-Mg-Al alloy system and is known to consist of six coordination shells having icosahedral symmetry for the T phase [5,8,9]. Icosahedral quasicrystals are also classified into P- and F-type quasicrystals [4,10–12]. One difference found between them experimentally is the scaling law with respect to the arrangement of reflections in electron diffraction patterns with the beam incidence parallel to the twofold axis. The arrangement of reflections in P-type quasicrystals exhibits $\tau^3$ scaling, while $\tau$ scaling is observed for F-type quasicrystals, where $\tau$ is a golden number.

The F-type icosahedral quasicrystal involving the Bergman cluster was found in the Zn-Mg-Y alloy system around the composition of 30at.%Mg-10at.%Y [11,13]. According to the reported phase diagram of the Zn-Mg-Y alloy system, there are some intermetallic-compound phases such as the H, W, and Z phases around this composition [14]. In addition, the $Zn_{23}Y_6$ phase is also present in the binary alloy system [15]. Because the icosahedral quasicrystal in the Zn-Mg-Y alloy system is expected to be stable around the valence-electron concentration of $e/a \sim 2.10$ for $Zn_{60}Mg_{30}Y_{10}$ in terms of the Hume-Rothery mechanism [16], in this study, we are interested in the crystallographic relationship among the H phase, $Zn_{23}Y_6$ phase, and the icosahedral quasicrystal, where $e/a$ values of the H ($Zn_{75}Mg_{10}Y_{15}$) and $Zn_{23}Y_6$ phases are estimated to be 2.15 and about 2.21, respectively. The crystal structures of the H and $Zn_{23}Y_6$ phases can commonly be regarded as coordination-polyhedra structures in metals and alloys such as the Laves, α-Mn, and β-Samson structures [15,17–26]. It is known that coordination-polyhedra structures consist of complex coordination polyhedra with higher coordination numbers (CNs) such as CN12 and CN16 polyhedra. Based on the reported crystallographic data, the $Zn_{23}Y_6$ structure involves CN8, CN12, CN13, and CN16 polyhedra, while CN12, CN15, and CN17 polyhedra are present in the H structure [15,17].

As mentioned above, the Bergman cluster for the T phase in the Zn-Mg-Al alloy system consists of six coordination shells with icosahedral symmetry [5,9]. Concretely, the first, second, third, fourth, fifth, and sixth shells have already been analysed to be the icosahedron, dodecahedron, icosahedron, truncated icosahedron, dodecahedron, and icosahedron, respectively. Because the icosahedron is just a CN12 polyhedron, the CN12 polyhedron must play an essential role in the formation of the Bergman cluster. Accordingly, the formation of the Bergman cluster from coordination-polyhedra structures with the CN12 polyhedron was investigated in this study, in particular, for the icosahedral quasicrystal in the Zn-Mg-Y alloy system, in which the H and $Zn_{23}Y_6$ phases with the CN12 polyhedron are present. Concretely, the crystallographic features of prepared Zn-Mg-Y samples with compositions of $Zn_{80-x+y}Mg_xY_{20-y}$ were examined mainly by transmission electron microscopy to clarify the crystallographic relations among the H structure, the $Zn_{23}Y_6$ structure, and the icosahedral quasicrystal. In the formula $Zn_{80-x+y}Mg_xY_{20-y}$, the compositions of the icosahedral-quasicrystal, the H phase, and the $Zn_{23}Y_6$ phase may be specified by (x = 30, y = 10), (x = 10, y = 5), and (x ~ 0, y ~ 0), respectively. As a result, it was found that there existed a simple crystallographic relationship among them. Based on the determined relationship, both the possible nucleus of the Bergman cluster and the features of its formation from the H and $Zn_{23}Y_6$ structures were also investigated in this study.

**II. Experimental procedure**

The present observation made by transmission electron microscopy was performed using $Zn_{80-x+y}Mg_xY_{20-y}$ alloy samples with various x and y values. Ingots of Zn-Mg-Y alloys were first prepared from Zn, Mg, and Y with purity of 99.9% by an induction-melting technique. For the purpose of homogenization, the alloy ingots were annealed at 623K for 24h, followed by quenching into ice water. It will be noted that no further thermal treatment was done to obtain possible stable and metastable states expected in $Zn_{80-x+y}Mg_xY_{20-y}$ alloy samples. To identify stable and metastable states appearing in the samples, their x-ray powder diffraction profiles were measured in the angular range of $10° \leq 2\theta \leq 100°$ at room temperature using a Rigaku SmartLab diffractometer with Cu$K\alpha$ radiation. The chemical composition of each region appearing in the samples was determined by means of a JSM-7001F-type scanning electron microscope equipped with an energy dispersive X-ray spectrometer (EDS). Their crystallographic features were also examined by taking electron diffraction patterns and corresponding bright- and dark-field images at room temperature using a JEM-3010-type transmission electron microscope with an accelerating voltage of 300 kV. Thin specimens for observation by transmission electron microscopy were prepared by an Ar-ion thinning technique.

**III. Experimental results**

In this study, x-ray powder diffraction profiles of $Zn_{80-x+y}Mg_xY_{20-y}$ alloy samples were measured at room temperature to identify stable and metastable states appearing in them. It was found that there were the F-type icosahedral quasicrystal and the hexagonal-$P6_3/mmc$ H phase in the samples, including the possible presence of the cubic-$F\bar{m}3m$ $Zn_{23}Y_6$ phase. The observation made by transmission electron microscopy indicated that the $Zn_{23}Y_6$ phase was in fact present as minor regions in a sample with the nominal composition of (x = 12, y = 6). We then tried to determine the crystallographic relations between the F-type icosahedral quasicrystal and these two intermetallic-compound phases. Although sample rotations of about 1° were required to obtain a simple relationship, we were able to establish the simple crystallographic relations between the F-type icosahedral quasicrystal and the H structure and also between the quasicrystal and the $Zn_{23}Y_6$ structure. The experimentally obtained data on the crystallographic relations will be presented below.

Figure 1 shows an x-ray powder diffraction profile in the angular range of $30° \leq 2\theta \leq 50°$, which was measured from an alloy sample with the nominal composition of (x = 12, y = 6) at room temperature, together with its bright-field images and corresponding electron diffraction patterns. According to the reported cross-section of the Zn-Mg-Al phase diagram at 700 K, this composition is located in the interior of the single H state [14]. As shown in Fig. 1(a), the measured profile was found to be very complicated, and reflections with stronger intensities were concentrated on around $2\theta = 38°$. From a comparison with the previously reported profiles of the icosahedral quasicrystal and the intermetallic-compound phases found in the Zn-Mg-Y alloy system, it was confirmed that the sample mainly consisted of icosahedral-quasicrystal and H-structure regions, whose reflections are indicated by the blue and yellow arrows in the profile, respectively. The point to note here is that some of the H reflections can also be indexed in terms of the $Zn_{23}Y_6$ structure with a lattice parameter of $a$ = 1.269 nm, as marked by the green arrows. To confirm the presence of these states, electron diffraction patterns and corresponding bright- and dark-field images were taken from various regions in the sample at room temperature by transmission electron microscopy. The electron diffraction patterns taken from two different areas in the same sample are, respectively, shown in Figs. 1(b) and 1(b') and 1(c) and 1(c'), together with their bright-field images in the insets. From these four patterns, large H and icosahedral-quasicrystal regions giving rise to relatively uniform contrasts in the images are confirmed to be present in the (x = 12, y = 6) sample. A simple analysis of the patterns indicated that the electron beam incidences of the two patterns in Fig. 1(b) and 1(b') are, respectively, parallel to the $N(2\bar{1}\bar{1}0)_H$ and $N(0001)_H$ directions in the H structure, where $N(hkml)_H$ means the normal direction of the $(hkml)_H$ plane. On the other hand, the patterns in Fig. 1(c) and (c') have the beam incidences parallel to the fivefold and twofold axes in the quasicrystal. It will be noted that τ scaling for the arrangement of reflections is also confirmed along the fivefold axis in Fig. 1(c'), as indicated by the small arrows. This is an experimental indication that the quasicrystal present in the sample was probably the F-type icosahedral quasicrystal. As suggested by the analysis of the x-ray powder diffraction profile, $Zn_{23}Y_6$ regions were also detected as minor regions, although experimental data on the $Zn_{23}Y_6$ state are not presented here. The chemical compositions of the icosahedral-quasicrystal, H, and $Zn_{23}Y_6$ regions were, respectively, estimated to be (x = 15, y = 10), (x = 10, y = 4), and (x = 6, y = 0) on the basis of the EDS analysis results.

    The alloy sample with the nominal composition of (x = 12, y = 6) was found to consist of F-type icosahedral-quasicrystal, hexagonal-H, and cubic-$Zn_{23}Y_6$ regions. We tried to determine the orientation relations between the quasicrystal and the H structure and between the quasicrystal

and the $Zn_{23}Y_6$ structure. The orientation relation between the quasicrystal and the H structure was first determined by using their coexistence areas in the sample. Figure 2 shows an electron diffraction pattern obtained from one of the coexistence areas including a quasicrystal/H boundary, together with its corresponding bright-field image in the inset. The electron beam incidence of the pattern is nearly parallel to the fivefold axis in the quasicrystal. As seen in the pattern, the fivefold-axis direction in the quasicrystal for the electron beam incidence is nearly parallel to the $N(2\bar{1}\bar{1}0)_H$ direction in the H structure. A sample rotation of about 1.5° was required to obtain the coincidence between the fivefold-axis and $N(2\bar{1}\bar{1}0)_H$ directions. The twofold-axis direction in the former is also parallel to the $N(05\bar{5}3)_H$ direction in the latter, as indicated by the yellow arrow. It is thus understood that, although lattice rotation of about 1.5° is taken into account, the orientation relationship of $N(2\bar{1}\bar{1}0)_H$ // the fivefold axis and $N(05\bar{5}3)_H$ // the twofold axis can be obtained between the F-type icosahedral quasicrystal and the hexagonal-H structure.

As mentioned above, the $Zn_{23}Y_6$ state was present as minor regions in the sample. We also determined the orientation relation between the icosahedral quasicrystal and the $Zn_{23}Y_6$ structure by using their coexistence areas. Figure 3 shows electron diffraction patterns and a corresponding bright-field image, which were taken from one of the coexistence areas including a quasicrystal/$Zn_{23}Y_6$ boundary. We first look at the image of the coexistence area in the inset, which was used for the analysis. It is seen that there are bright- and dark-contrast regions separated by a curved boundary, which are referred to as Regions A and B. To identify these regions, we took their electron diffraction patterns with various electron beam incidences at room temperature. The pattern in Fig. 3(a) was taken from Region A, while two patterns from Region B are shown in Figs. 3(b) and 3(b'). The pattern of Region A exhibits an arrangement of reflections with fivefold symmetry with respect to the origin 000. From an analysis of electron diffraction patterns with various beam incidences, including the pattern in Fig. 3(a), the state in Region A was identified as the F-type icosahedral quasicrystal, just as in the case of the above-mentioned (quasicrystal + H) coexistence area. On the other hand, the patterns of Region B show a simple regular arrangement of reflections, as seen in Figs. 3(b) and 3(b'). To determine the crystal structure in Region B, we constructed its reciprocal lattice by using electron diffraction patterns obtained experimentally. The constructed reciprocal lattice for Region B is schematically depicted in the inset in Fig. 3(b'). Based on the determined reciprocal lattice, it was found that the state in Region B was entirely consistent with the $Zn_{23}Y_6$ state with cubic-$Fm\bar{3}m$ symmetry.

The orientation relationship between the F-type icosahedral quasicrystal and the $Zn_{23}Y_6$ structure was determined by taking electron diffraction patterns of the above-mentioned area including a boundary between neighbouring icosahedral-quasicrystal and $Zn_{23}Y_6$ regions. Figure 4 shows one of the superposed patterns for these two regions, together with a schematic diagram. A very complex pattern is observed because of the involvement of the two regions. It will be noted that an electron beam incidence for one of the two regions deviates slightly from a higher-symmetry direction for the other. Concretely, when the $Zn_{23}Y_6$ region with the $[110]_c$ incidence was rotated by an angle of about 0.6° about the $[\bar{1}13]_c$ direction, the direction parallel to the twofold axis in the quasicrystal was obtained. The schematic diagram of the superposed pattern was then depicted by taking the 0.6°-rotation into account, where the reflections due to the icosahedral quasicrystal and the $Zn_{23}Y_6$ structure are represented by the red and black closed circles, respectively. In the schematic diagram, the $[110]_c$ beam incidence for the cubic-$Zn_{23}Y_6$ structure is parallel to the twofold axis in the quasicrystal. The location of the $\bar{1}13_c$ reflection in the former is also coincident with that of the $11\bar{1}\bar{1}13$ reflection in the latter, as indicated by the small open circle. This implies that the $[\bar{1}13]_c$ direction in the $Zn_{23}Y_6$ structure is parallel to one of the fivefold axes in the quasicrystal, where the fivefold-axis direction is indicated by the thick arrow in the schematic diagram. Another important point to note here is that the $11\bar{1}\bar{1}13$ reflection in the quasicrystal is one of the characteristic features in reciprocal space for the F-type icosahedral quasicrystal. That is, the orientation relation of $(\bar{1}13)_c$ // the fivefold axis and $[110]_c$ // the twofold axis was established between the cubic-$Fm\bar{3}m$ $Zn_{23}Y_6$ structure and the F-type icosahedral quasicrystal, although a slight rotation of about 0.6° was required to obtain the relation. In the alloy sample with the nominal composition of (x = 12, y = 6), we thus found the orientation relationship among the cubic-$Fm\bar{3}m$ $Zn_{23}Y_6$ structure, the F-type icosahedral quasicrystal (IQ), and the hexagonal-$P6_3/mmc$ H structure; $[\bar{1}13]_c$ // the fivefold axis in the IQ // $N(2\bar{1}\bar{1}0)_H$, and $[110]_c$ // the twofold axis in the IQ // $N(05\bar{5}3)_H$, although slight rotations of about 1° were needed to obtain this relationship.

**IV. Discussion**

From our experimentally obtained data mentioned above, it was confirmed that there were F-type icosahedral-quasicrystal, H, and $Zn_{23}Y_6$ regions, for instance, in the alloy sample with the nominal composition of (x = 12, y = 6). An interesting feature of the samples used in this study is that $Zn_{23}Y_6$ regions were also present as minor regions. We also found the crystallographic relationship of $[\bar{1}13]_c$ // the fivefold axis in the IQ// $N(2\bar{1}\bar{1}0)_H$, and $[110]_c$ // the twofold axis in the IQ// $N(05\bar{5}3)_H$, although a slight rotation of about 1° was needed to obtain the relationship. The simple relation provided information for understanding the formation of the Bergman cluster in the F-type icosahedral quasicrystal from the $Zn_{23}Y_6$ and H structures. Based on the determined relationship, we discuss here both a possible nucleus of the Bergman cluster and its formation from the $Zn_{23}Y_6$ and H structures.

The discussion starts with the correspondence between atomic positions in the $Zn_{23}Y_6$ structure and in the Bergman cluster involved in the F-type icosahedral quasicrystal. Concretely, we made an assignment of a region in the $Zn_{23}Y_6$ structure, corresponding to four inner shells: that is, the first, second, third, and fourth shells in the Bergman cluster. The reason for adopting four shells is that the lattice parameter of $a = 1.269$ nm in the cubic-$Zn_{23}Y_6$ structure is longer than the diameter of about 0.986 nm for the third shell in the case of the approximant-T structure in the Al-Mg-Zn alloy system and is slightly shorter than that of about 1.356 nm for the fourth shell. In addition, the 116 atoms involved in the unit cell of the $Zn_{23}Y_6$ structure are comparable to the 104 atoms of the four inner shells. With the help of the determined crystallographic relation, we first looked for a CN12 polyhedron in the $Zn_{23}Y_6$ structure, which has the same orientation as a CN12 polyhedron located at the centre of the Bergman cluster. Figure 5 shows superposed projections of atoms in both the $Zn_{23}Y_6$ structure and the four inner shells of the Bergman cluster, together with schematic diagrams indicating the correspondence between atomic positions in the $Zn_{23}Y_6$ structure and those in the Bergman cluster for each inner shell. In the projections and the diagrams, the grey and brown circles represent Zn and Y atoms in the $Zn_{23}Y_6$ structure, while atoms in the first, second, third, and fourth shells are indicated by the red, blue, green, and yellow circles. The unit cell of the $Zn_{23}Y_6$ structure and the four shells in the Bergman cluster are also indicated by the black lines, and the red, blue, green, and yellow lines, respectively. In addition, the projected directions in Figs. 5(a) and 5(b) are, respectively, parallel to the $[110]_c$ and $[001]_c$ directions in the $Zn_{23}Y_6$ structure, and the crystallographic data reported by Kuz'ma, Kripyakevich, and Frankevich were used to determine the locations of atoms in the $Zn_{23}Y_6$ structure in this study [15]. From these two projections, it was found that the projected atomic positions in the $Zn_{23}Y_6$ structure and those in the four inner shells of the Bergman cluster resembled each other. This is an experimental indication that there is basically a one-to-one correspondence between atomic positions in the $Zn_{23}Y_6$ structure and those in the four inner shells. Based on this one-to-one correspondence, the positional correspondence between atoms in the $Zn_{23}Y_6$ structure and those in each shell is shown in Fig. 5(c). Atomic shifts needed for the formation of each shell from the $Zn_{23}Y_6$ structure are also indicated by the black and red arrows. As seen in the diagrams, the first and second shells of the Bergman cluster can be formed from the $Zn_{23}Y_6$ structure by relatively small shifts. On the other hand, large atomic shifts are required for some atoms in the third and fourth shells, as indicated by the red arrows. We then discuss the origin of the large atomic shifts in the formation of the third shell. The atoms with large shifts are found on the shell of a CN16 polyhedron, whose centre atoms are large Y atoms in the $Zn_{23}Y_6$ structure for the first shell, as indicated by the brown arrows. This implies that large atomic shifts can be avoided by an atomic replacement of a centre atom from the larger Y atom to a small Zn one. Apparently, this replacement produces a coordination-number change of (CN16 → CN12). In other words, the first, second, and third shells in the Bergman cluster can be formed only by the atomic replacement of four Y atoms in the first shell to four Zn atoms in the second shell for the $Zn_{23}Y_6$ structure. As a result, among the four inner shells, twelve Zn atoms with CN12 form the icosahedron as the first shell, while the dodecahedron as the second shell consists of twelve Zn atoms and eight Y atoms.

In the inner shells produced from the $Zn_{23}Y_6$ structure by the atomic replacement, eight Y atoms with CN16 in the second shell are also required to occupy special positions to produce the first, second, and third shells in the Bergman cluster. The required positions of the eight Y atoms in the second shell are schematically depicted in Fig. 6, together with a structural block consisting of eight CN16 polyhedra. In Fig. 6(a), the eight required positions in the second shell are indicated by the large blue circles. The interesting features of the eight required positions are that a cube can be obtained by connecting two neighbouring positions, and that each side of the cube is equal to a diagonal line of a pentagon, which is present on the surface of the dodecahedron as the second shell. Furthermore, we can also construct the structure block indicated by the blue colour in Fig. 6(b) by putting together eight CN16 polyhedra with a planar contact, whose centre Y atoms sit on the eight required positions. The striking point to note here is that an arrangement of atoms in the interior of the structural block forms the first, second, and third shells in the Bergman cluster. This is thus an indication that the structural block consisting of eight CN16 polyhedra in Fig. 6(b) may be a possible nucleus of the Bergman cluster.

  Another structural block as a possible nucleus of the Bergman cluster can also be expressed in terms of six short-penetrated-decagonal columns. Figure 7 shows schematic diagrams indicating a continuous change between these two possible expressions for the nucleus of the Bergman cluster. The starting expression for the nucleus is assumed to be the structural block consisting of eight CN16 polyhedra. In the starting block, eight CN16 polyhedra are referred to as Polyhedra 1, 2, 3, 4, 5, 6, 7, and 8, as indicated in diagram I, where Polyhedron 7 is not seen. After taking out Polyhedra 1, 2, 3, and 4, we obtain diagram II, where there is a CN12 polyhedron at the centre position of the structural block, as marked by the red lines. As shown in diagram III, next, we can find a short-penetrated-decagonal column indicated by the pink lines. The features of the penetrated-decagonal column are that it consists of three CN12 polyhedra, and that its axis is parallel to one of six fivefold axes in the structural block. In the block, six short-penetrated-decagonal columns are also present along the six fivefold axes, with the common CN12 polyhedron at the centre. As a result, we can depict another structural block marked by the red color in diagram IV, which is characterized by an asembly of six short-penetrated-decagonal columns. In other words, this structural block as an assembly of six short-decagonal columns is another possible candidate for the nucleus of the Bergman cluster, in addition to the structural block consisting of eight CN16 polyhedra.

  Keeping the above-mentioned discussion in mind, the correspondence between atomic positions in the H structure and those in the Bergman cluster is discussed on the basis of the determined crystallographic relationship. The orientation relation between the H structure and the F-type icosahedral quasicrystal can be given by $N(2\bar{1}\bar{1}0)_H$ // the fivefold axis in the IQ and $N(055\bar{3})_H$ // the twofold axis in the IQ, although sample rotation of about 1.5° was needed to obtain it. Based on this determined relation, we superposed atomic positions in both the H structure and the inner four shells in the Bergman cluster so that a centre CN12 polyhedron in the Bergman cluster coincided with one of the CN12 polyhedra in the H structure. Figures 8(a) and 8(b) show two projections of superposed atomic positions along the $N(2\bar{1}\bar{1}0)_H$ and $N(0001)_H$ directions in the H-structure notation, respectively. In the figure, silver, brown, and silver-brown circles represent Zn, Y, and mixed-site atoms in the H structure, respectively, while atoms in the first, second, third, and fourth shells in the Bergman cluster are indicated by the smaller red, blue, green, and yellow circles. In addition, the first, second, third, and fourth shells in the cluster are also depicted by the red, blue, green, and yellow lines, and the crystallographic data on the H structure reported by Deng *et al.* were used to obtain their atomic positions. Although it is hard to see the positional correspondence at first glance, the one-to-one correspondence can be found between atomic positions in the H structure and those in the Bergman cluster. A diagram indicating the positional correspondence between atoms in the H structure and those in each of the four inner shells is shown in Fig. 8(c). Atomic shifts for the formation of each shell from the H structure are also indicated by the black and red arrows. As seen in the diagram, atomic positions in each shell can be obtained by very simple atomic shifts. Among these shifts, those indicated by the small red arrows for the third shell seem to be somewhat larger than the others. A simple analysis of the positional correspondence indicated that the larger shifts originate from the presence of two Y atoms in the H structure for the first shell. We can thus avoid the relatively larger shifts by the exchange of two Y atoms in the first shell to two Zn atoms in the second shell. In other words, atoms forming the first, second, and third shells in the Bergman cluster are potentially present in the intermetallic-compound H structure.

The positional correspondence between atoms in the $Zn_{23}Y_6$ structure and those in the Bergman cluster suggested that there are two possible candidates for the nucleus in the formation of the Bergman cluster from the $Zn_{23}Y_6$ structure; that is, structural block I consisting of eight CN16 polyhedra and block II as an asembly of six short-penetrated-decagonal columns. To understand which candidate is more appropriate, the locations of atoms in the H structure for the first and second shells in the Bergman cluster are checked in terms of these two structural blocks. In Fig. 8(c), we first focus on the twenty atoms in the H structure, which form the second shell of the Bergman cluster. From the schematic diagram for the second shell, the corresponding atoms in the H structure consist of six Zn atoms with CN12, six Y atoms with CN17, and eight mixed-site atoms with CN12 or CN15. For the second shell there is no atom with CN16 in the H structure. This implies that the twenty atoms in the H structure for the second shell cannot serve as a centre atom of a CN16 polyhedron for structural block I. This means that structural block I based on eight CN16 polyhedra should be ruled out as a possible nucleus. Among twelve atoms in the H structure for the first shell, on the other hand, both six Zn atoms and four mixed-site atoms can be identified as centre atoms of a CN12 polyhedron. As a matter of fact, four short-penetrated-decagonal columns can be found in the H structure, with the common CN12 polyhedron at the centre. When the other two Y atoms in the first shell are replaced by two Zn atoms in the second shell, structural block II can be constructed as an asembly of six short-penetrated-decagonal columns in the H structure. Based on this, we can conclude that an asembly of six short-penetrated-decagonal columns indicated by the red colour in diagram IV in Fig. 7 is the most appropriate nucleus of the Bergman cluster in its formation from the $Zn_{23}Y_6$ and H structures.

Finally, we simply discuss the formation of the Bergman cluster from the determined nucleus consisting of the first, second, and third shells. As mentioned above, an asembly of six short-penetrated-decagonal columns as the nucleus of the Bergman cluster can be formed from the H structure by a simple atomic replacement of two Y atoms in the first shell to two Zn atoms in the second shell. Because the Bergman cluster involves six shells, for the approximant-T structure [5,9], the nucleus must grow to form the fourth, fifth, and sixth shells. To promote this growth, we conducted some trials by using the determined nucleus of the Bergman cluster. As a result, it was found that the formation of the fourth and sixth shells, except for the fifth shell, could be achieved by the adoption of a short-penetrated-decagonal column consisting of five CN12 polyhedra, not three CN12 polyhedra. To reproduce the fifth shell in the Bergman cluster, all twenty positions in the second shell must be occupied by 16 coordination atoms such as Y and Mg atoms. This means that the formation of the Bergman cluster cannot be explained in terms of only the development of an asembly of six short-penetrated-decagonal columns as the possible nucleus. The occupation of 16 coordination atoms in the second shell is another important factor for the formation of the Bergman cluster.

**V. Conclusions**

The crystallographic relations between the F-type icosahedral quasicrystal involving the Bergman cluster and the intermetallic-compound phases in the Zn-Mg-Y alloy system were examined mainly by transmission electron microscopy. It was found that there existed a simple crystallographic relationship among the cubic-$Fm\bar{3}m$ $Zn_{23}Y_6$ structure, the hexagonal-$P6_3/mmc$ H structure, and the F-type icosahedral quasicrystal (IQ); that is, $[\bar{1}13]_c$ // the fivefold axis in the IQ // $N(2\bar{1}\bar{1}0)_H$, and $[110]_c$ // the twofold axis in the IQ // $N(05\bar{5}3)_H$. Both the nucleus of the Bergman cluster and its formation from the $Zn_{23}Y_6$ and H structures were then discussed on the basis of the experimentally obtained relationship. Based on that relationship, an asembly of six short-penetrated-decagonal columns was identified as an appropriate nucleus of the Bergman cluster, which involves the first, second, and third shells. The growth of the nucleus to the Bergman cluster was explained in terms of the following two factors; the development of an asembly of short-penetrated-decagonal columns as the nucleus and the occupation of 16 coordination atoms such as Y and Mg atoms in the second shell. This is an indication that the H phase may be identified as one of the approximant phases for the F-type icosahedral quasicrystal.

Figure captions

Figure 1. X-ray powder diffraction profile of a Zn-Mg-Y alloy sample with the nominal composition of (x = 12, y = 6) for the formula $Zn_{80-x+y}Mg_xY_{20-y}$, together with its bright-field images and corresponding electron diffraction patterns. The profile in (a) was measured at room temperature in the angular range of $30° \leq 2\theta \leq 50°$. The reflections in the profile are basically explained as being due to the icosahedral quasicrystal and the intermetallic-compound H phase, as indicated by the blue and yellow arrows. It was also found that some of the H reflections may be indexed in terms of the $Zn_{23}Y_6$ structure, as marked by the green arrows. An observation made by transmission electron microscopy confirmed that the sample consisted of two major regions and one minor region. Of two major regions, the pattern and the image for one region are shown in (b), (b'), and the inset of (b), while those in (c), (c'), and the inset of (c) were obtained from the other. Based on these patterns, the states of two major regions were identified as the H phase for the former region and the F-type icosahedral quasicrystal for the latter. The F-type quasicrystal was confirmed by an array of reflections indicated by the arrows in (c').

Figure 2. Electron diffraction pattern and a corresponding bright-field image of the sample with the nominal (x = 12, y = 6) composition at room temperature, obtained from one of the coexistence areas including a boundary between the icosahedral-quasicrystal and H regions. The electron beam incidence of the pattern is nearly parallel to the fivefold axis in the quasicrystal and to the $N(2\bar{1}\bar{1}0)_H$ direction in the H structure. A sample rotation of about 1.5° was required to obtain the coincidence between the fivefold-axis and $N(2\bar{1}\bar{1}0)_H$ directions. As indicated by the yellow arrow, the twofold-axis direction in the quasicrystal is also parallel to the $N(05\bar{5}3)_H$ direction in the H structure.

Figure 3. Electron diffraction patterns and a corresponding bright-field image of the sample with the nominal (x = 12, y = 6) composition at room temperature, taken from one of the coexistence areas including a quasicrystal/$Zn_{23}Y_6$ boundary. Two regions separated by a curved boundary are present in the area shown in the inset in (a) for the image and are referred to as Regions A and B. Region A gives rise to the pattern in (a), while two patterns in (b) and (b') were taken from Region B. Based on the arrangement of reflections with fivefold symmetry in (a), the state of Region A is identified as the icosahedral quasicrystal. On the other hand, the patterns of Region B show a simple regular arrangement of reflections. From the determined reciprocal lattice depicted in the inset in (b'), the state of Region B is understood to be entirely consistent with the $Zn_{23}Y_6$ state with cubic-$Fm\bar{3}m$ symmetry.

Figure 4. Superposed electron diffraction patterns obtained from the area including neighbouring icosahedral-quasicrystal and $Zn_{23}Y_6$ regions in Fig. 3, together with a schematic diagram. The electron beam incidence of the pattern is nearly parallel to the twofold axis in the quasicrystal and to the $[110]_c$ direction in the $Zn_{23}Y_6$ structure. Rotation of the $[110]_c$ incidence in the $Zn_{23}Y_6$ structure by an angle of about 0.6° was required to obtain the twofold axis in the quasicrystal. The construction of the diagram takes into account the rotation of about 0.6°, and the reflections due to the icosahedral quasicrystal and the $Zn_{23}Y_6$ structure are represented by the red and black closed circles, respectively. The location of the $113_c$ reflection in the $Zn_{23}Y_6$ structure is coincident with that of the $11\bar{1}\bar{1}13$ reflection in the quasicrystal, as indicated by the small open circle. This implies that the $[\bar{1}13]_c$ direction in the $Zn_{23}Y_6$ structure is parallel to one of the fivefold axes in the quasicrystal.

Figure 5. Superposed projections of atoms in both the $Zn_{23}Y_6$ structure and the four inner shells of the Bergman cluster, together with schematic diagrams indicating the positional correspondence between atoms in the $Zn_{23}Y_6$ structure and those in the Bergman cluster for each shell. Apparently, the four inner shells involve the first, second, third, and four shells. The projections were made along the $[110]_c$ direction for (a) and the $[001]_c$ one for (b) in the $Zn_{23}Y_6$-structure notation. In the figure, Zn and Y atoms in the $Zn_{23}Y_6$ structure, and atoms in the first, second, third, and fourth shells in the Bergman cluster are, respectively, represented by the grey and brown, and small red, blue, green, and yellow circles. The unit cell of the $Zn_{23}Y_6$ structure is also indicated by the black lines. In (c), atomic shifts needed for the formation of each shell from the $Zn_{23}Y_6$ structure are indicated by the black and red arrows. It was found that the large atomic shifts indicated by the red arrows in the third shell originated from the presence of large Y atoms marked by the brown arrows in the $Zn_{23}Y_6$ structure for the first shell.

Figure 6. Schematic diagrams showing the required positions of eight Y atoms in the second shell, together with a structural block consisting of eight CN16 polyhedra. The required positions indicated by the large blue circles in (a) are needed to form the first, second, and third shells in the Bergman cluster. The structure block shown in (b) can be constructed by positioning eight CN16 polyhedra with centre Y atoms, which sit on the required positions. The structural block indicated

by the blue colour, which consists of eight CN16 polyhedra, may be a possible nucleus of the Bergman cluster.

Figure 7. Schematic diagrams showing a continuous change between two candidates for the possible nucleus of the Bergman cluster. In diagram I, the structural block consisting of eight CN16 polyhedra is first shown as a starting nucleus. Among eight CN16 polyhedra referred to as Polyhedra 1, 2, 3, 4, 5, 6, 7, and 8 in diagram I, Polyhedra 1, 2, 3, and 4 are removed to create diagrams II and III. A CN12 polyhedron and a short-penetrated-decagonal column are seen in the interior of the structural block, as indicated by the red lines in diagram II and the pink lines in diagram III. In the structural block, six short-penetrated-decagonal columns are present along the six fivefold axes, with a common CN12 polyhedron at the centre. As a result, a structural block as an assembly of six short-decagonal columns can be obtained as the other candidate, as marked by the red color in diagram IV.

Figure 8. Superposed projections of atoms in both the H structure and the four inner shells of the Bergman cluster, together with diagrams indicating the positional correspondence between atoms in the H structure and those in the Bergman cluster for each shell. The projections were made along the $N(2\bar{1}\bar{1}0)_H$ direction for (a) and the $N(0001)_H$ one for (b) in the H-structure notation. In the figure, the silver, brown, and silver-brown circles represent Zn, Y, and mixed-site atoms in the H structure, while atoms in the first, second, third, and fourth shells in the Bergman cluster are indicated by the small red, blue, green, and yellow circles. In (c), atomic shifts for the formation of each shell from the H structure are also indicated by the black and red arrows. The relatively large atomic shifts indicated by the red arrows in the third shell were found to originate from the presence of two large Y atoms marked by the brown arrows in the H structure for the first shell.

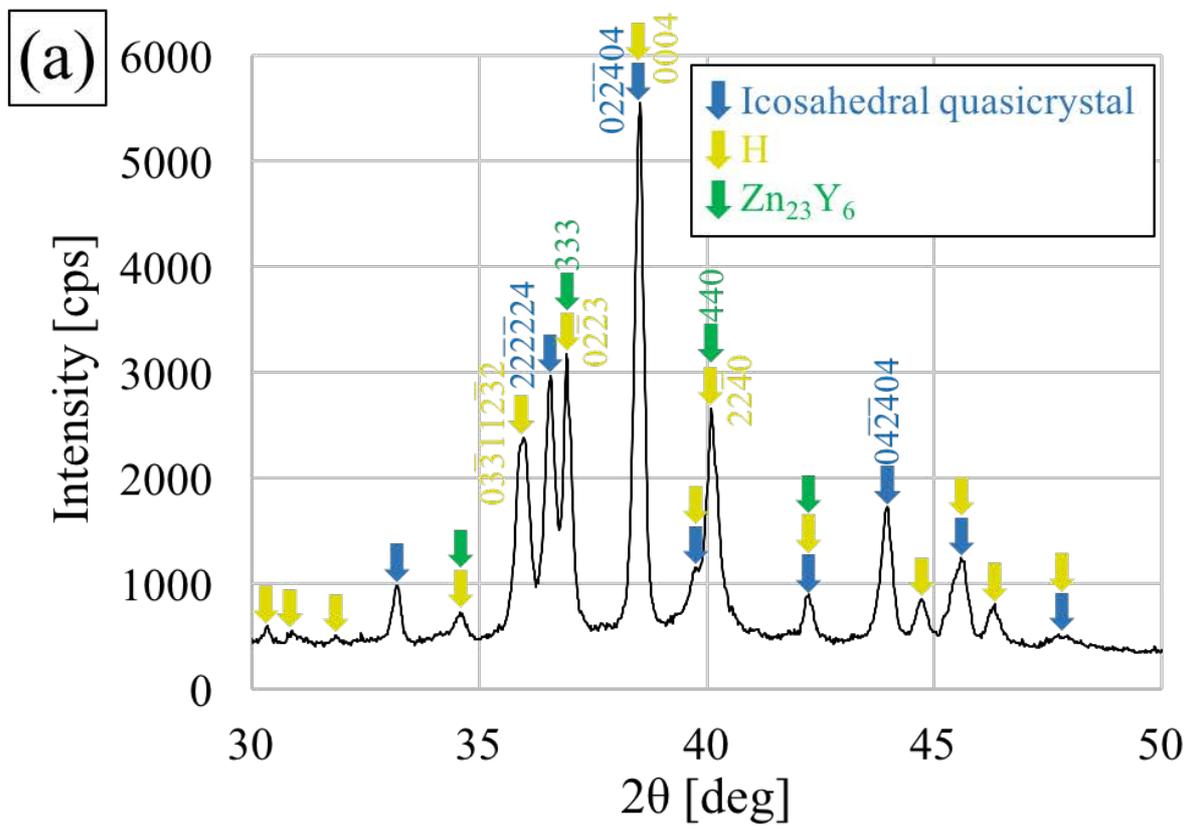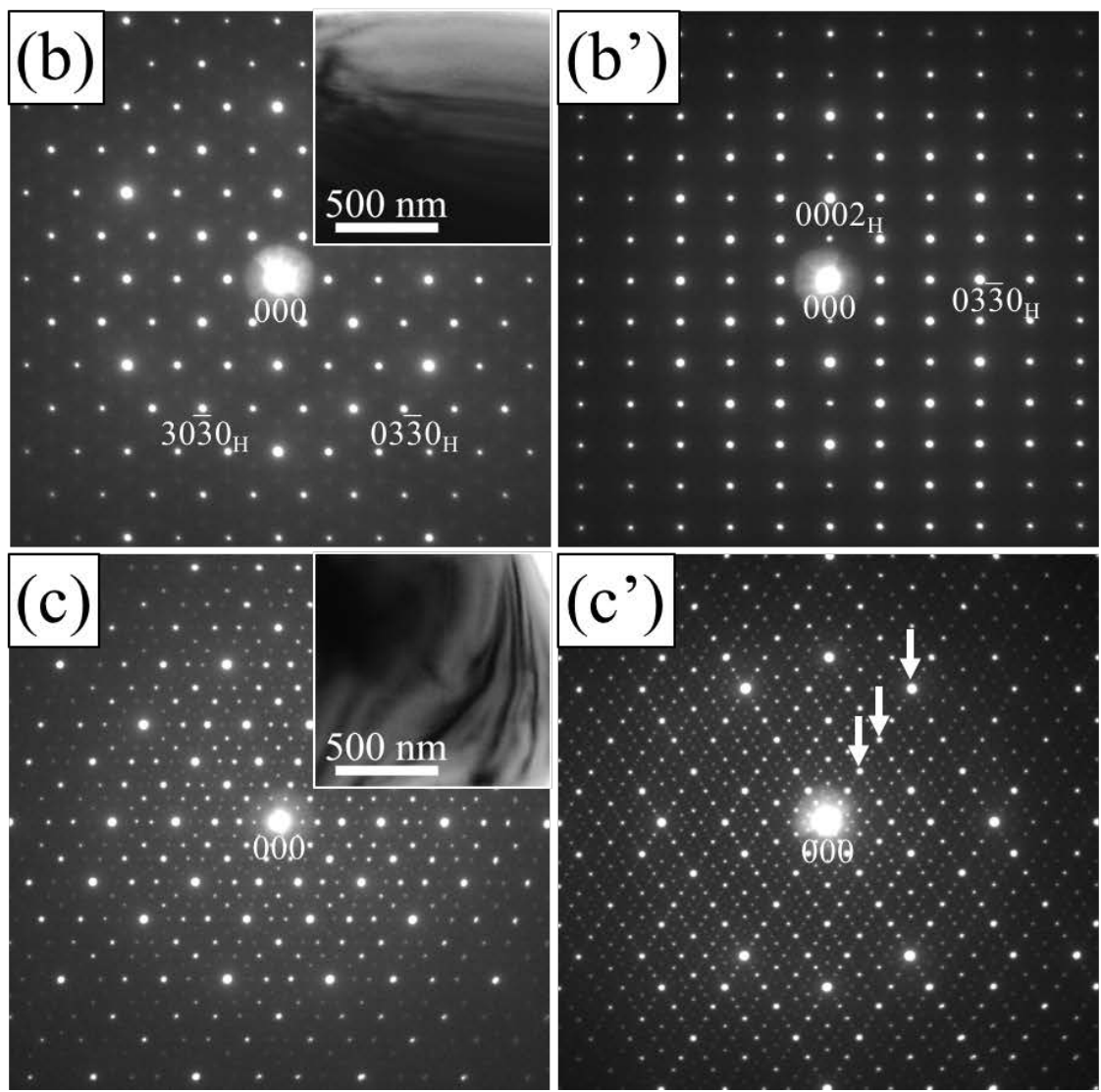

Fig. 1

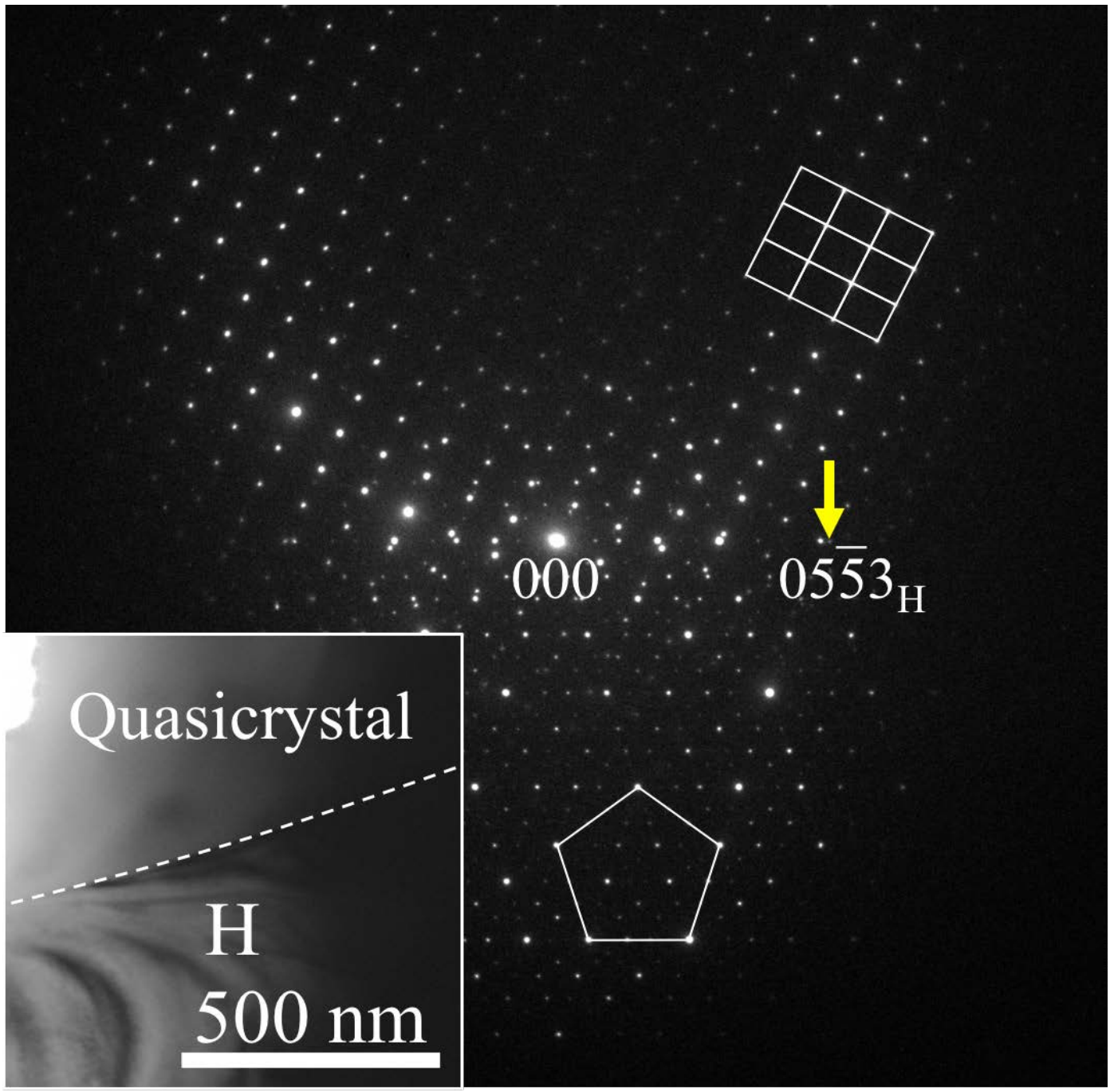

Fig. 2

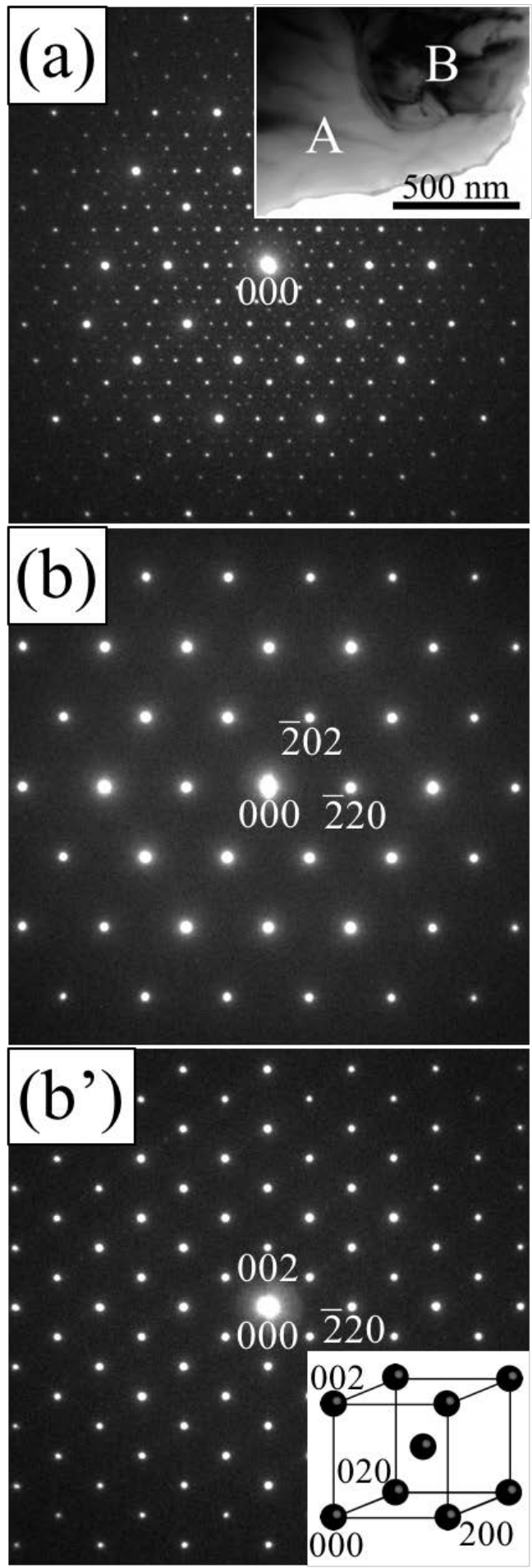

Fig. 3

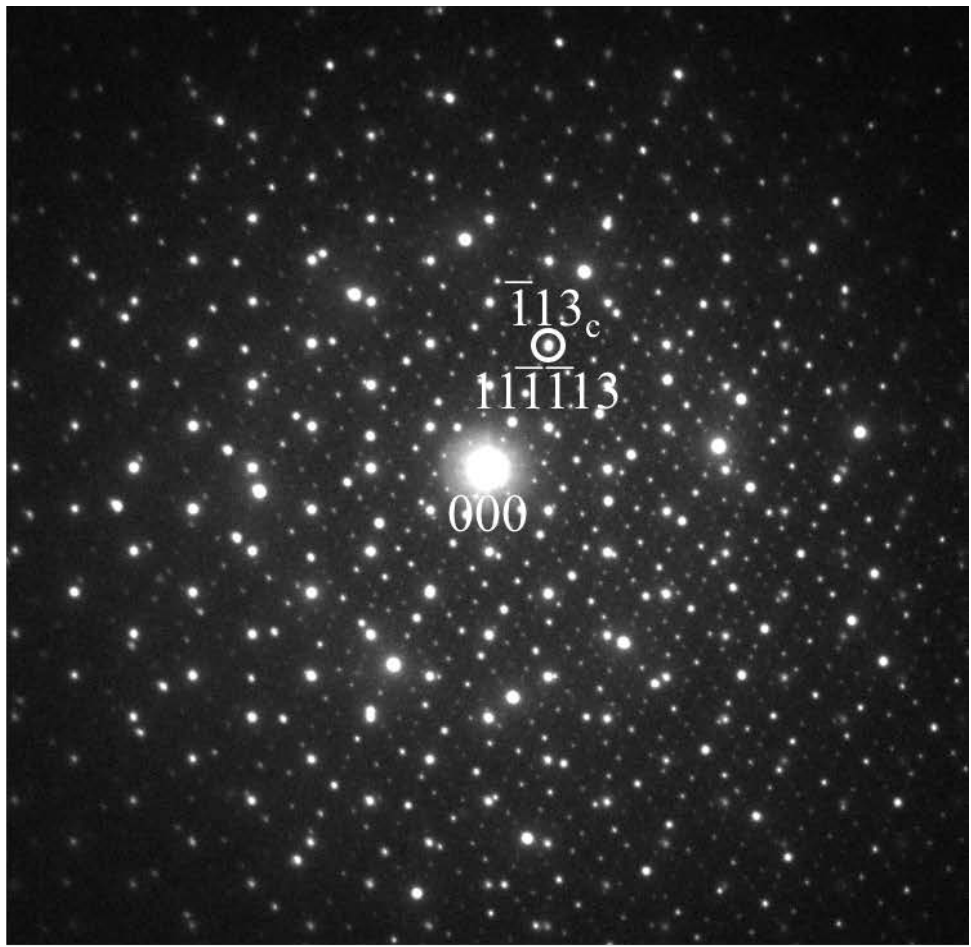

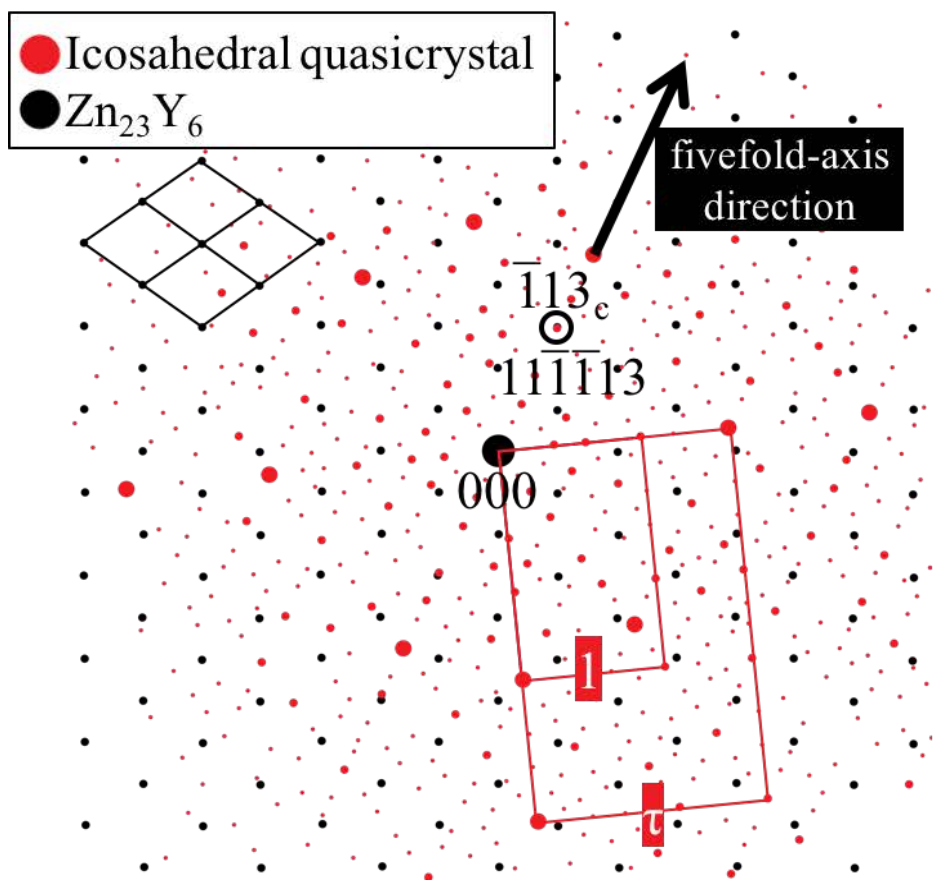

Fig. 4

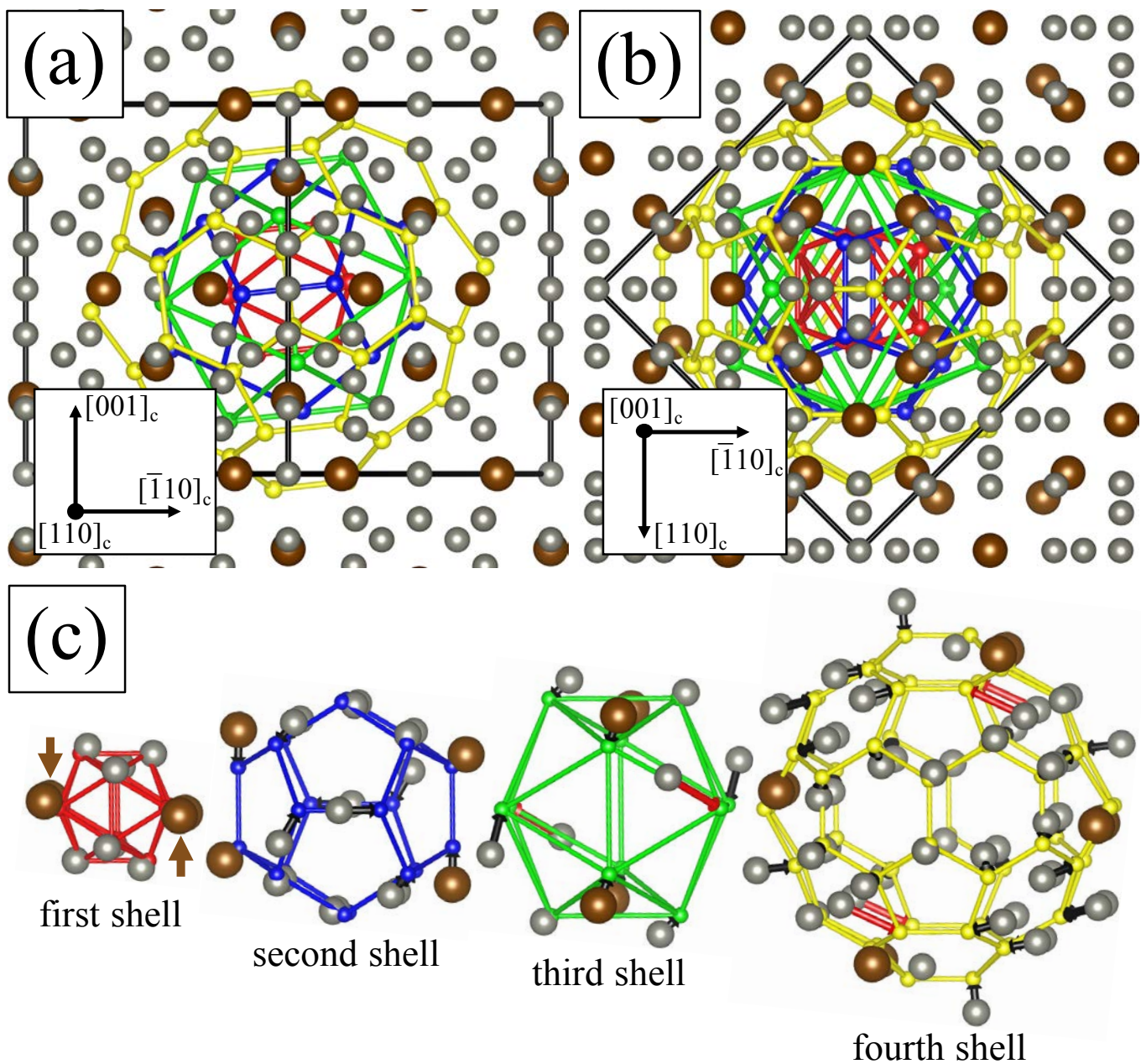

Fig. 5

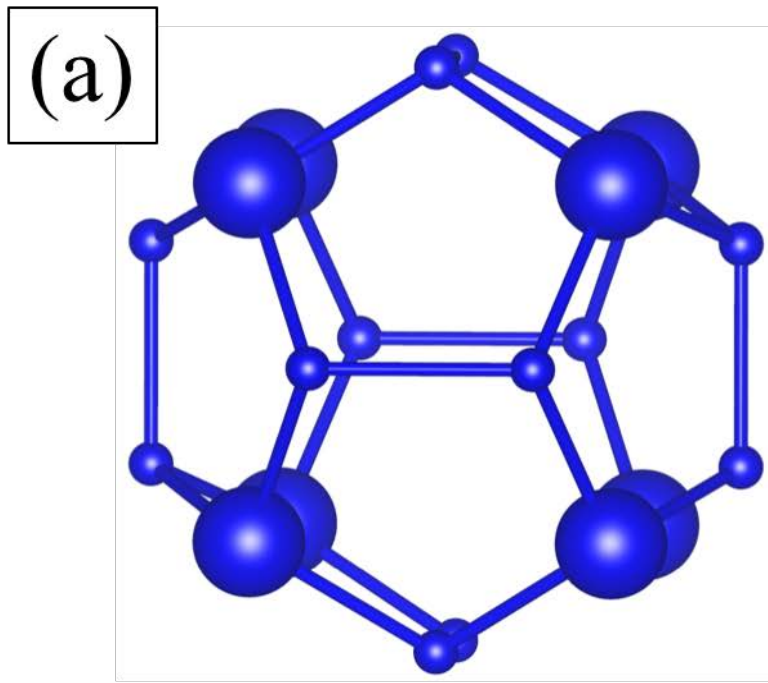

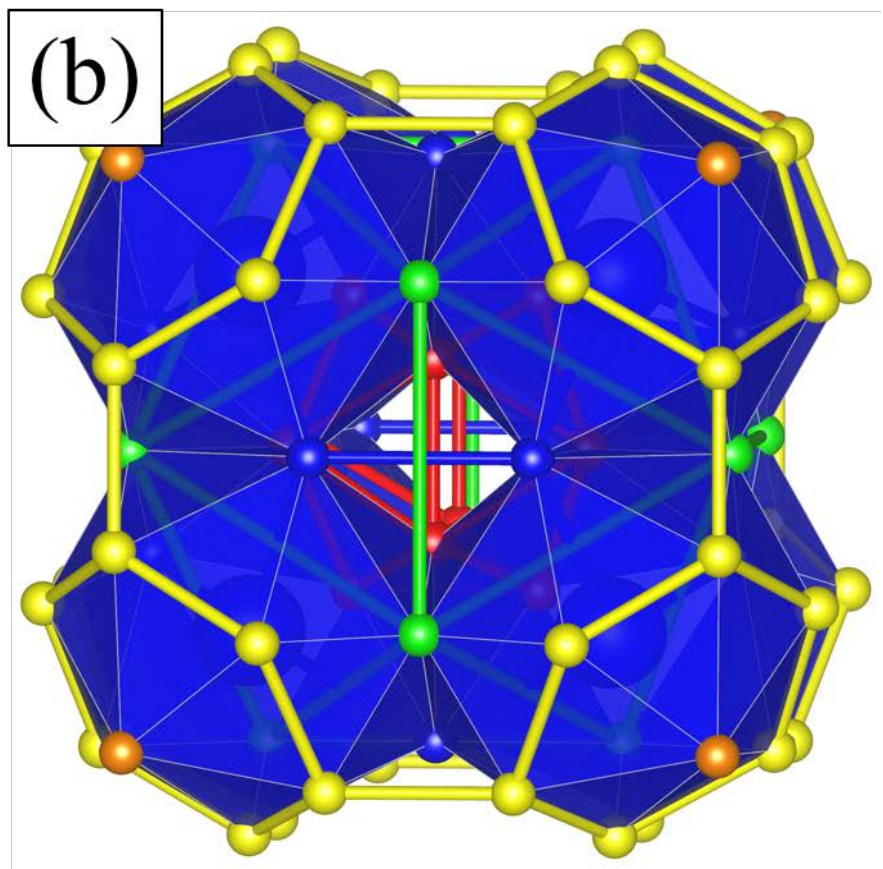

Fig. 6

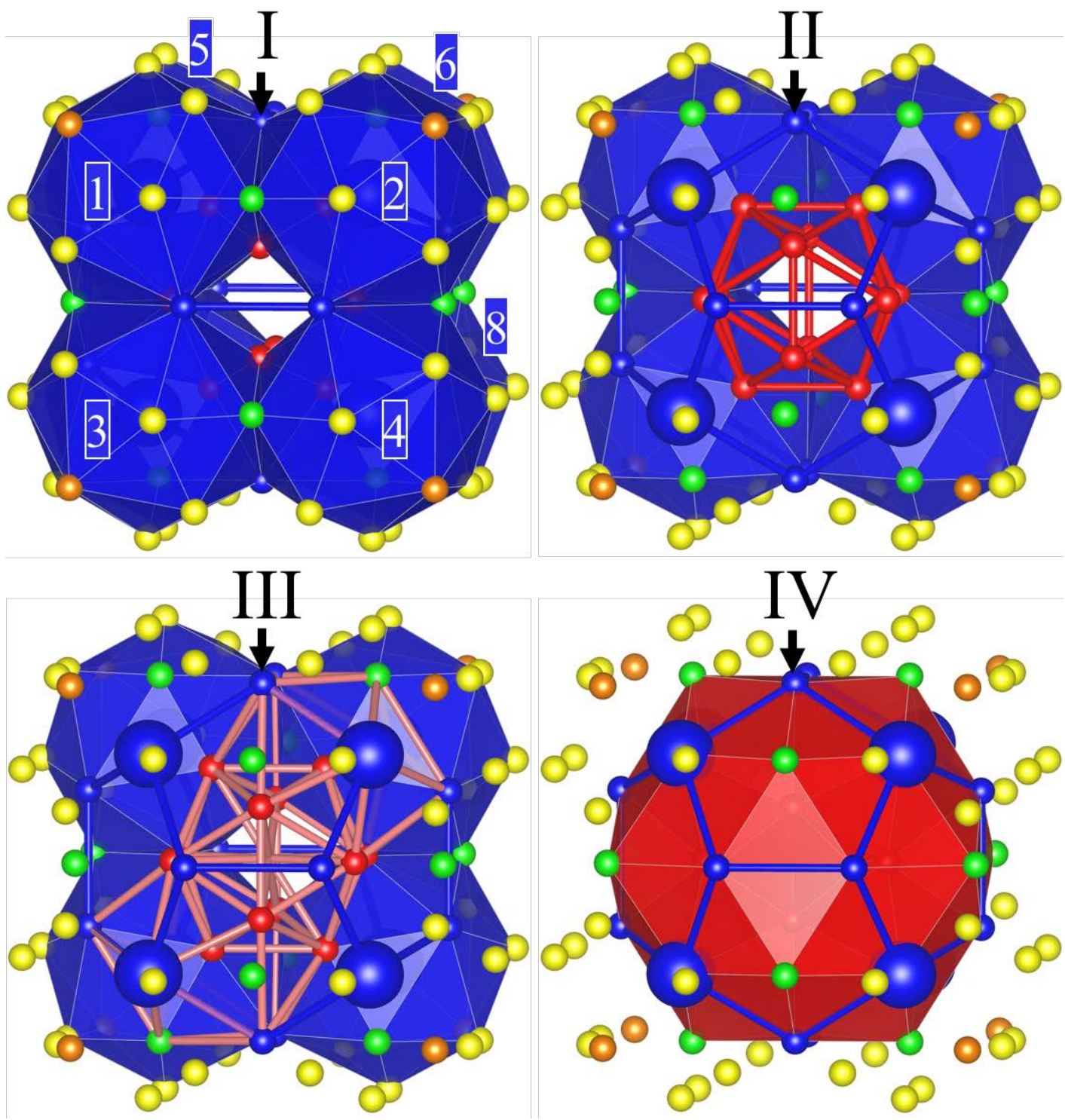

Fig. 7

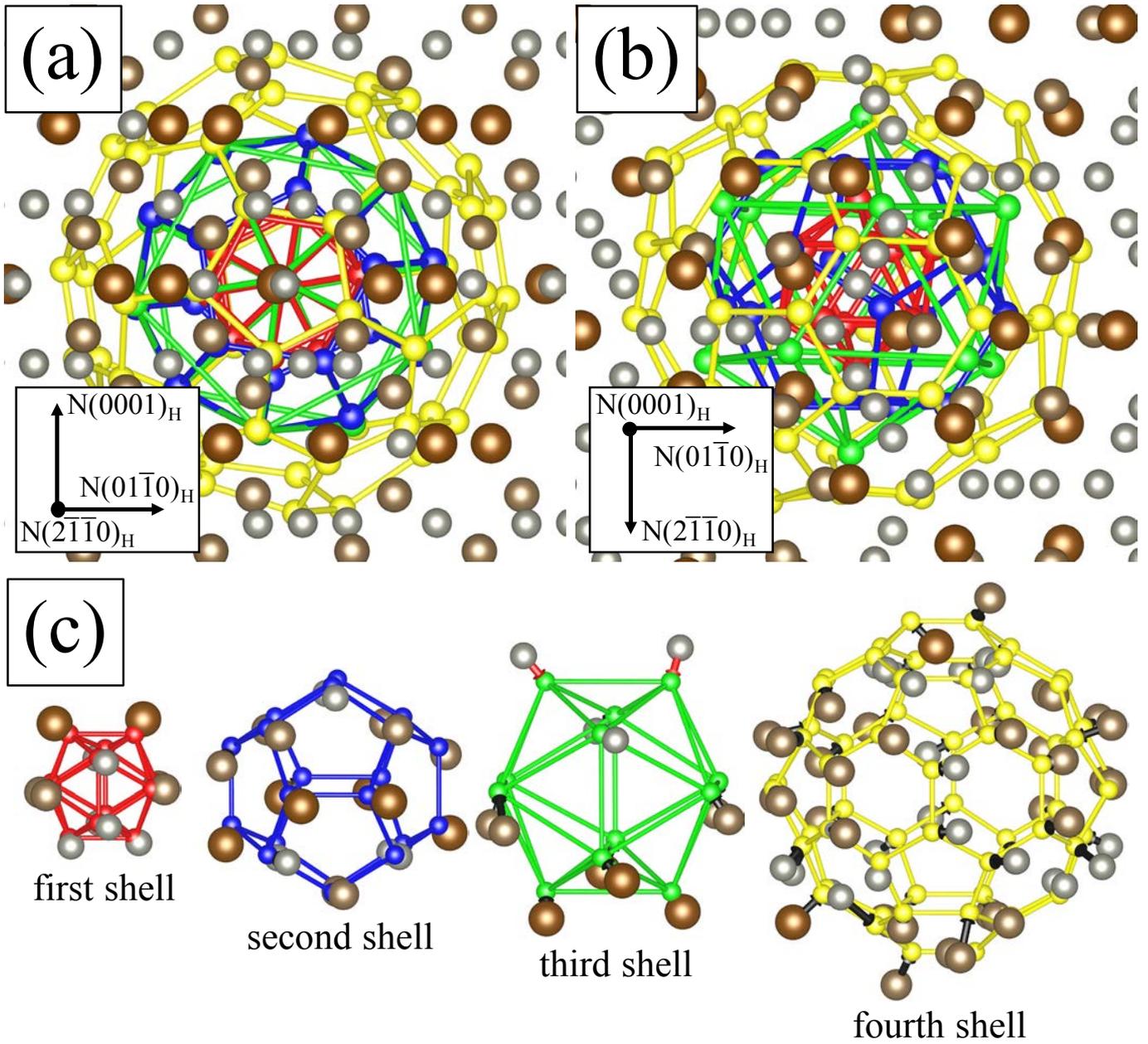

Fig. 8